\documentclass[11.5pt,a4,epsf]{article}

\usepackage{amssymb,amsmath,algorithm}
\usepackage{xcolor}

\usepackage{graphicx}
\usepackage{rotating}
\usepackage{float}
\usepackage{amssymb,amsmath,algorithm}

\usepackage{float}

\usepackage{graphicx}
\DeclareGraphicsExtensions{.eps}
\usepackage{epsfig}
\usepackage{epstopdf}
\usepackage{amsmath}
\usepackage{amssymb}
\usepackage{subfig}
\usepackage{rotating}
\usepackage{wrapfig}
\usepackage{makeidx,epsfig}
\usepackage[margin=1.0in]{geometry}
\usepackage{amssymb,amsfonts}
\usepackage{amsmath}
\usepackage{amsbsy}
\usepackage{verbatim}
\usepackage[bottom]{footmisc}
\usepackage{algorithm}
\usepackage{algorithmicx}
\usepackage[noend]{algpseudocode}
\algdef{SE}{Begin}{End}{\textbf{begin}}{\textbf{end}}
\usepackage{colortbl,xcolor,pgfplots}
\usepackage{algorithm,algorithmicx,algpseudocode}
\usepackage{xcolor}
\usepackage{makeidx}
\usepackage{capt-of}
\usepackage[T1]{fontenc}
\usepackage{amsfonts}
\usepackage{amscd}
\usepackage{amssymb}
\usepackage{tabularx}
\usepackage{amssymb,bezier,graphicx,}
\usepackage{times,amssymb}
\usepackage{graphicx}
\usepackage{color}
\usepackage{cancel}
\usepackage{fancybox}
\usepackage{fancyhdr}
\usepackage{hyperref}
\usepackage{tikz}
\usepackage{enumerate}
\usepackage{epstopdf}
\usepackage{array}
\usepackage{slashbox}
\usepackage{verbatim}
\usepackage{float}
\usepackage[colorinlistoftodos]{todonotes}
\usepackage{framed}
\usepackage{anysize}
\usepackage{pdfpages}
\usepackage{stackrel}
\usepackage{booktabs}
\usepackage[mathscr]{eucal}
\usepackage{bm}
\usepackage{multirow}
\usepackage{caption}
\usepackage{mathrsfs} 
\usepackage{xcolor}
\usepackage[normalem]{ulem}
\usepackage{marginnote}

 \newtheorem{definition}{\bf Definition}

\begin{document}

\title{Evaluating conditional covariance estimates via a new targeting approach and a networks-based analysis.}

\author{Carlo Drago$^1$, Andrea Scozzari$^{1,2}$\\
\\
\normalsize $^1$ Universit\`{a} degli Studi Niccol\`{o} Cusano Roma, \normalsize  carlo.drago@unicusano.it\\
\normalsize $^2$ CESDE, Centro Studi per l'Analisi delle Dinamiche Economiche\\ Universit\`{a} Niccol\`{o} Cusano, \normalsize  andrea.scozzari@unicusano.it}

\maketitle


\begin{abstract}
Modeling and forecasting of dynamically varying covariances have received much attention in the literature. The two most widely used conditional covariances and correlations models are BEKK and DCC. In this paper, we advance a new method to introduce targeting in both models to estimate matrices associated with financial time series. Our approach is based on specific groups of highly correlated assets in a financial market, and these relationships remain unaltered over time. Based on the estimated parameters, we evaluate our targeting method on simulated series by referring to two well-known loss functions introduced in the literature and Network analysis. We find all the maximal cliques in correlation graphs to evaluate the effectiveness of our method. Results from an empirical case study are encouraging, mainly when the number of assets is not large.

\end{abstract}

\section{Introduction}
Modeling financial volatility is a crucial issue in empirical finance. \emph{Volatility clustering} is a peculiar feature of a financial time series, that is, the series is characterized by periods more volatile than others due to unpredictable outer events, or, in other words, the series suffers from \emph{heteroscedasticity}. The existence of heteroscedasticity is a major concern in a univariate approach aiming at estimating and forecasting a given (financial) phenomenon. In 1982 Robert Engle \cite{Engle_1982} introduced formal methods of analyzing economic time series with time-varying volatility named the Auto Regressive Conditional Heteroscedasticity (ARCH) model. This model was further generalized by Bollerslev in \cite{Bollerslev_1986} and renamed as the Generalized ARCH (GARCH) model. Modeling financial volatility becomes a more difficult task if we consider $N$ time series (multivariate approach) and we are interested in understanding the co-movements of financial returns. It is therefore important to extend the analysis to multivariate GARCH (MGARCH) models. In fact, these models play an important role in different fields of finance like, for instance, in portfolio optimization \cite{EngCola_2006}, option pricing \cite{RomSteVio_2014}, analysis of contagion and volatility spillovers \cite{BilCapo_2010,ChaLiMcA_2019}.
In the literature MGARCH models are classified into different categories (see, e.g., \cite{BauLauRom_2006}): (1) direct generalizations of the univariate GARCH models; (2) linear combinations of univariate GARCH models, such as, generalized orthogonal models and latent factor models; (3) nonlinear combinations of univariate GARCH models; (4) nonparametric and semiparametric models that constitute an alternative to the parametric estimation of the financial volatility that do not impose a particular structure on the data. In this paper we will focus on categories (1) and (3).
\vskip 8 pt
\noindent In finance, the main aim of multivariate GARCH models is to predict the future values of the variance and covariance matrix of asset returns. The literature has received much attention for modeling and forecasting dynamically varying covariances, with numerous multivariate ARCH, GARCH, and stochastic volatility specifications proposed for this problem. Research on these models has been lively ever since 1986. Readers may refer to the very recent paper \cite{SilTer_2021} and the references for a comprehensive review of the existing literature on MGARCH approaches.

Modeling a covariance matrix is difficult because of the likely high dimensionality of the problem and the constraint that a covariance matrix must be positive definite. The first developed MGARCH models of category (1) were the \emph{VEC} model introduced in \cite{BollEngWool_1988}, and the \emph{BEKK} one proposed in \cite{EngleKron_1995}. Both approaches deal with the direct estimation of the covariance matrix between the assets returns. One disadvantage of the VEC model is that there is no guarantee of a positive semi-definite estimated matrix. On the other hand, the BEKK model has the attractive property that the conditional covariance matrices are positive definite by construction. However, estimation of a BEKK model involves somewhat heavy computations due to it contains a large number of parameters. To overcome this problem, one approach was to decompose the conditional covariance matrix into conditional standard deviations and a conditional correlation matrix (models of category (3)). In the first model proposed in \cite{Bollerslev_1990} the conditional correlation is assumed to be constant over time, and only the conditional standard deviation is time-varying, hence the name Constant Conditional Correlation (\emph{CCC-GARCH}) model. The restriction of constant correlations over time was not considered always reasonable and was later removed. In \cite{EngShep_2001} and \cite{Engle_2002} the authors defined the \emph{Dynamic} Conditional Correlation model (\emph{DCC-GARCH}), which is an extension of the CCC-GARCH one, for which the conditional correlation matrix is designed to vary over time. This approach has one important advantage, that is, the number of parameters to be simultaneously estimated is reduced since a complex optimization problem is disaggregated into simpler ones.

\vskip 8 pt
\noindent Several authors provided a number of different modifications of conditional covariance MGARCH models, some of them are mentioned in \cite{SilTer_2021}. Other meaningful approaches explicitly incorporates the effect of measurement errors and time-varying attenuation biases into the covariance estimation and forecasts (see, e.g., \cite{BollPatQua_2018}). Engle and Kelly \cite{EngKel_2012} studied a special case of the DCC-GARCH model, the dynamic \emph{equicorrelation} model, in which all correlations are identical. This can be a useful model in situations where the number of parameters to estimate are large.

All the MGARCH models presented in the literature estimate the conditional covariance or correlation matrices via the optimization of an appropriate log-likelihood function. Once having obtained the parameters, the following step consists in an estimation evaluation phase, which is performed with several methods. In particular, when comparing different approaches, the estimations are usually compared w.r.t. some measures like the Mean Absolute Error (MAE) and/or the Mean Square Error (MSE) (see, e.g., \cite{Patton_2011}). Alternatively, the performance of the estimated conditional covariance and correlation matrices is measured by using some \emph{loss} functions that measure the distance from the fitted covariance matrix from the ex-post realized one. Among others we mention the Frobenius distance, the quasi-likelihood measure, and the Kullback-Leibler divergence \cite{BollPatQua_2018,LauRomVio_2013}. Finally, in order to determine the goodness of fit of the residuals, univariate and/or multivariate approaches consisting of the application of a number of different tests are performed, such as, for example the Ljung-Box test.

\vskip 8 pt
\noindent In this paper we advance a novel approach based on network analysis for evaluating the estimates of the time-varying correlation matrices in financial markets. First, we consider the two most widely used MGARCH models, that is, the BEKK and DCC ones, and propose a variant of these two models by suitably modifying the log-likelihood function to maximize. We call \emph{modified BEKK} and \emph{modified DCC} the two resulting models, respectively. This modification consists in introducing a term in the function incorporating a loss measure based on the difference between time-varying covariance matrices and the covariance matrix estimated w.r.t. the whole in-sample period. In fact, as observed in \cite{StoRacRachFab_2011}, a financial market is characterized by some stylized facts. In particular, there are often specific groups of assets that are highly correlated in such a way that, for instance, positive price changes of one asset in the group determine positive price changes of all the other assets in the group, and these relationships remain unaltered over time. Hence, the idea beyond our modification is that (extremely) high/low values of correlations observed between pairs of assets do not change too much during time, more precisely, any two highly positive or negative correlated assets in the market remain highly (positive/negative) correlated during the given observed time period.
\\
The correlation among assets in a financial market can be pointed out via graph theory in which, given a time period, assets are identified by vertices of a complete graph $G$ and distances (weights) assigned to pairs of assets (edges) incorporate the dependency structure of returns. Mantegna \cite{Mantegna_1999} was one of the first to construct asset graphs based on stocks prices correlations in order to detect the hierarchical organization inside a stock market. In order to highlight only groups of assets whose correlation is above or below a given threshold $\delta$, one can extract a (sub)graph $G(\delta)$ of $G$ that contains only a subset of the connections in $G$ between pairs of assets. In the new graph $G(\delta)$ two stocks are linked if and only if they exhibit high/low values of correlation. In other words, the \emph{observed} subgraph $G(\delta)$ reveals the strong relationships between groups of assets in a financial market during a given time period $T$. Examples of such interconnected groups are, for instance, firms belonging to a specific industrial sector.\\
\\
In this paper, the additional term introduced in the log-likelihood function takes into account that the time-varying covariance or correlation matrices must not alter the clusters formed by highly correlated assets observed during $T$. Hence, with our method we estimate a modified MGARCH model, simulate $T$ realizations of returns of each of the $N$ assets, compute the correlation matrix w.r.t. the whole simulated in-sample period and, given the threshold $\delta$, we obtain the corresponding \emph{simulated} graph $G_S(\delta)$. Then, we compare the observed and simulated graphs $G(\delta)$ and $G_S(\delta)$, respectively. In particular, we compare all the \emph{maximal Cliques} of the two graphs. Given a general graph $G$, a \emph{Clique} is a set of interconnected stocks that forms a complete subgraph of $G$. Hence, Cliques in $G(\delta)$ and in $G_S(\delta)$ represent highly (positive/negative) correlated assets, that is, any stock that belongs to a clique is highly correlated with all other stocks in the clique. The comparisons allow us verifying if a modified MGARCH model has been able to correctly reproduce the volatility and all the strong relationships among assets of the given financial market, and if a modified model is able to outperform the corresponding original MGARCH model.

\subsection{Motivation}

A similar modification of the two above considered models through the introduction of a \emph{target} matrix was provided in \cite{CapMcA_2012}. In that paper, the authors directly modified the ``structure'' of the BEKK and DCC models with the target matrix and compare their modified models with the aim of improving estimate of the conditional covariance and correlation matrices. The authors consider the long-run covariance matrix as a target matrix, which is consistently estimated by the corresponding sample estimator, and consider different versions of the two MGARCH models. Finally, they showed that, from a theoretical perspective, the optimal model for estimating conditional covariances was the scalar BEKK one, regardless of whether targeting was used (see \cite{CapMcA_2012}). In particular, they point out that, other than for the scalar BEKK model, imposing positive definiteness and covariance stationarity for other considered versions of BEKK is extremely complicated when there is more than one asset. For the DCC models, targeting can be useful to reduce the number of parameters to estimate. However, in \cite{Aielli_2009} the authors proved that the estimation of DCC models with targeting may be inconsistent. Therefore, the suggestion is that one has to reconsider all models with targeting.\\
\\
\noindent In any case, from a theoretical viewpoint, the paper \cite{CapMcA_2012} is worth of interesting since the authors study the availability of analytical forms for the sufficient conditions for consistency and asymptotic normality of the appropriate estimators and computational tractability for large problem dimensions. From an empirical point of view, in \cite{CapMcA_2012} no empirical tests are presented, but the authors conclude that it would not seem to be possible to provide an appropriate judgment a priori as to which of the different MGARCH models is to be preferred.
\\
\\
In light of the above considerations, the contribution of the present paper is threefold. First, we propose a new alternative targeting method that does not alter the models' structure. In fact, we modify only the log-likelihood objective function by introducing a suitable loss or distance measure between the long-run sample covariance or correlation matrix and the corresponding conditional covariance or conditional correlation matrices, respectively. Hence, we do not have to impose additional constraints, for instance, for the covariance stationarity or for guaranteeing the matrices to be definite positive. Second, we provide an empirical analysis for evaluating our modified models. We do this with a recommendation that it is not our goal to definitely decide which of the different models is the best one. Rather, we seek to understand whether our modified models allow to better capture the strong relationships between assets in a stock market. At the very least, we hope that the modified models can be more accurate that the standard ones. Third, we advance a new method for evaluating the effectiveness of the estimated models through some specific tools commonly used in network analysis.
\\
\\
\noindent The paper is organized as follows: Section \ref{Sec:Properties} reports notation and some basic definitions. Sections \ref{Bekk} and \ref{DCC} resume the two MGARCH models considered in the paper along with the new log-likelihood functions used in the optimization/estimation phase. Section \ref{Data_analysis} reports the structure of the empirical financial dataset we employed in the experimental phase described in Section \ref{Test}. Finally, some conclusions and further research are depicted in Section \ref{Conclusions}

\section{Notation, definitions}
\label{Sec:Properties}

Consider a financial market formed by $N$ assets. Let $P_{jt}$ be the daily closing price of asset $j$ at time $t$, $t=0,\ldots,T$, and $r_{jt}=\log(\frac{P_{jt}}{P_{jt-1}})$ be the corresponding log-return of asset $j$. In the following, Pearson correlation coefficients are used to detect dependencies between assets returns. A general MGARCH model is defined:

\begin{equation}\label{GeneralGARCH}
\left\{
\begin{array}{lll}
\bf{r_t}&=&\bf{\mu_t + \epsilon_t}
\\
\bf{\epsilon_t}&=&\bf{H^{\frac{1}{2}}_t \eta_t},
\end{array}
\right.
\end{equation}

\noindent where $\bf{r_t}$ is the $N\times 1$ vector of log-returns at time $t$, $\bf{\epsilon_t}$ is the $N\times 1$ vector of mean-corrected returns of $N$ assets at time $t$, with E[$\bf{\epsilon_t}$]=0 and Cov[$\bf{\epsilon_t}$] = $\bf{H_t}$. The vector $\bf{\mu_t}$ represents the expected value of $\bf{r_t}$. Observe that $\bf{\mu_t}$ may be modeled as a constant vector or as a time series model. In this paper we assume $\bf{\mu_t}$ constant.
\\
$\bf{H_t}$ is the $N\times N$ matrix of conditional variances and covariances of the unpredictable component $\bf{\epsilon_t}$ at time $t$. Finally, $\bf{\eta_t}$ is the $N\times 1$ vector of i.i.d errors such that E[$\bf{\eta_t}$]=0 and E[$\bf{\eta_t}$$\bf{\eta_t}'$]=$I$.\\
\\
In a general MGARCH model, we have to estimate the conditional covariance matrix $\bf{H_t}$, which, in addition, has to be positive definite for all $t$. Depending on the possible specifications for $\bf{H_t}$, we have different MGARCH models each belonging to one of the four categories mentioned before.
\vskip 8 pt

\noindent In order to estimate the conditional covariance matrix $\bf{H_t}$, a common issue using multivariate data is to resort to the maximization of an appropriate log-likelihood function $L(\bf{\theta})$, where $\bf{\theta}$ denotes the vector of all the parameters to estimate. Depending on the MGARCH model, the function $L(\bf{\theta})$ assumes a different appearance. It is well-known that the quality of the Maximum Likelihood (ML) estimation relies also on the assumed
data distribution. In general, when dealing with models with conditional heteroscedasticity, the estimates are known to be asymptotically normal \cite{BollWool_1992}. In our approach, we consider a multivariate Gaussian distribution for the standardized error $\bf{\eta_t}$, even if, in principle, our method can be applied assuming different distributions.
\vskip 8 pt
\noindent Given a stock market, consider $\bf{H_t}$ the conditional variance and covariance matrix of the returns of the $N$ assets at time $t$, and let $\bf{R}=[\rho_{ij}]$, with $\rho_{ii}=1$, $i=1,\ldots,N$, be the \emph{global} correlation matrix w.r.t. the sample period. The idea is that if asset $i$ is highly positively/negatively correlated to asset $j$, then this kind of (high positive/negative) correlation does not change during the observed period. Note that we assume that the relationship based on the correlation between two assets $i$ and $j$ does not change during time while the value of the correlation coefficient $\rho_{ij}$, can (obviously) vary.\\
\\
Let $\delta > 0$ be a threshold, we construct the correlation graph $G(\delta)=(V,E)$, where $V$ is the set of vertices each representing an asset, and $E$ is the set of edges of $G(\delta)$, that is, the set of connections between pairs of assets. We assume that there is an edge $(i,j)$  between assets $i$ and $j$ if and only if the (global) correlation coefficient $|\rho_{ij}|>\delta$. Let $A(\delta)$ be the adjacency matrix of $G(\delta)$. Observe that the generic element $a_{ij}(\delta)$ can be 0 or 1 if we assume an unweighted graph, otherwise  $a_{ij}(\delta)=w_{ij}$ if we assume that the graph $G(\delta)$ is weighted with weights $w_{ij}$ assigned to each edge $(i,j)$ and obtained as a function of the correlation coefficients $\rho_{ij}$, that is, $w_{ij}=f(\rho_{ij})$. Given a general graph $G=(V,E)$, the following definitions hold \cite{BonMur_2008, Christofides_1975}:

\begin{definition}
A subset $C\subseteq V$ is called a clique of $G$ if any two vertices in $C$ are connected by an edge. The order $q$ of a clique is the cardinality of $C$.
\end{definition}

\begin{definition}
A subset $C\subseteq V$ is called a maximal clique of order $q$ of $G$ if $C$ is not included in any other clique $C'$ of order $q+1$.
\end{definition}

\noindent Other concepts not defined in the paper can be found in the book \cite{BonMur_2008}.

\vskip 8 pt
\noindent In a modified MGARCH model, the \emph{modified} likelihood function to maximize is:

\begin{equation}\label{modified_Likelihooh}
L(\bf{\theta}) - \sum\limits_{t=1}^T [Dist(A(\delta), A_t(\delta)],
\end{equation}

\noindent where $Dist(\cdot)$ is (any) distance or divergence measure between the adjacency matrix $A(\delta)$ referred to the global correlation matrix $\bf{R}$ and the adjacency matrix $A_t(\delta)$ related to the conditional correlation matrix $\bf{R_t}$. Since $A(\delta)$ is the correlation matrix computed w.r.t. the sample period, it can be considered as a target matrix.

In our framework, as a measure we can consider any distance or divergence measure between two $N\times N$ positive definite matrices $P$ and $Q$. In this paper, we consider the well-known Kullback-Leibler (KL) distance (or divergence) \cite{KullLei_1951} between $P$ and $Q$, namely $KL(P,Q)$. As a statistical measure, it is a measure of the distance between two probability densities $P$ and $Q$. In the case of Gaussian multivariate distributions this distance is completely defined by the correlation matrices of the whole system. Thus, it can be interpreted as how a multivariate probability distribution represented by the matrix $Q$ is different from a reference multivariate probability distribution represented by the matrix $P$. It is also well-known that it is not a metric since this measure is not symmetric and does not satisfy the triangle inequality. Given matrices $P$ and $Q$, the $KL(P,Q)$ measure is:

\begin{equation}\label{KL_distance}
KL(P,Q)=\frac{1}{2}\big[ \log(\frac{|Q|}{|P|}) + Tr(Q^{-1}P) - N\big],
\end{equation}

\noindent where the operator $|\cdot|$ computes the determinant of a matrix and $Tr(\cdot)$ is the trace of a square matrix, which is the sum of its diagonal elements. In our approach $Dist(A(\delta), A_t(\delta))=KL(A(\delta), A_t(\delta))$.\\
\\
Note that assuming the identity function as the function of the graph's weights, i.e.  $w_{ij}=\rho_{ij}$, $i\ne j$, the matrix $A(\delta)$ corresponds, in fact, to the unconditional correlation matrix where, given $\delta$, $a_{ij}(\delta)=\rho_{ij}\ne 0$, $i\ne j$, if and only if $|\rho_{ij}| > \delta$, $a_{ij}(\delta)=0$, $i\ne j$, otherwise (i.e., $|\rho_{ij}| \leq \delta$), and further imposing $a_{ij}(\delta)=1$, $i=1,\ldots,N$. Denote by $\widehat{Z}$ the matrix $A(\delta)$ defined so far. Under the above hypothesis on the graph's weights, also $A_t(\delta)$ can be considered a conditional correlation matrix at time $t$. Hence, in this special case, in order to preserve the strong relationships among specific groups of assets, in the maximization of the modified likelihood function (\ref{modified_Likelihooh}) we are, in fact, requiring that $a_{ij}(\delta)$ and $a_{ij,t}(\delta)$ be as close as possible, that is, the corresponding correlation values $|\rho_{ij}|$ and $|\rho_{ij,t}|$ be as close as possible. In the rest of the paper we assume $w_{ij}=\rho_{ij}$, $i\ne j$.

\subsection{The BEKK model}\label{Bekk}

The BEKK class of multivariate GARCH models was introduced in \cite{EngleKron_1995}. In order to make the model easier to solve, Ding and Engle \cite{DingEngle_2001} proposed a variant of the original one namely the $\emph{diagonal}$ BEKK model. The general model is:

\begin{equation}\label{general_BEKK}
\bf{H_t}=C\cdot C'+\sum\limits_{k=1}^K\sum\limits_{i=1}^p A_{ki}\cdot (\epsilon_{t-i}\epsilon'_{t-i})\cdot A'_{ki} + \sum\limits_{k=1}^K\sum\limits_{j=1}^q B_{kj}\cdot H_{t-j}\cdot B'_{kj},
\end{equation}

\noindent where $A_{ki}$ and $B_{kj}$ are parameter matrices, $C$ is a lower triangular matrix, $p$ and $q$ represent the lagged error term and the number of conditional covariance lags, respectively. $K$ determines the generality of the process. We assume that $p=q=1$ and $K=1$, so that the diagonal BEKK model in a compact form can be written:

\begin{equation}\label{diagonal_BEKK}
\bf{H_t}=C\cdot C'+A\cdot (\epsilon_{t-1}\epsilon'_{t-1})\cdot A' + B\cdot H_{t-1}\cdot B',
\end{equation}

\noindent with $A$ and $B$ diagonal matrices. Positive definiteness of conditional covariance matrices is guaranteed, by
construction (see \cite{DingEngle_2001}). The procedure used in estimating the parameters of the model is the maximization of a likelihood function constructed under the assumption of an i.i.d. of the errors $\bf{\eta_t}$. Under the further assumption of a conditional normality, the set of all the parameters $\theta$ of the multivariate diagonal BEKK model can be estimated by maximizing the following sample log-likelihood function:

\begin{equation}\label{log-likelihood_Bekk}
L(\theta)=-\frac{TN}{2}\log(2\pi)-\frac{1}{2}\sum\limits_{t=1}^T \big(\log|H_t|+ \epsilon'_t H_{t}^{-1} \epsilon_t \big),
\end{equation}

\noindent with $T$ the number of returns observations. Note that in the BEKK model equation (\ref{diagonal_BEKK}) refers to the conditional covariance matrices. Thus, in the following modified log-likelihood function we have to use covariance matrices in place of the correlation matrices. Hence, under the assumption $w_{ij}=\rho_{ij}$, $i\ne j$, as a special case for the graph's weights, given $\delta$, we first compute matrix $\widehat{Z}$ and then the corresponding unconditional covariance matrix $\widehat{\Sigma}$ w.r.t. the whole sample period $T$, that is, $\widehat{\Sigma}=\Gamma\widehat{Z}\Gamma$, with $\Gamma$ the diagonal matrix of standard deviations w.r.t. the whole sample period. Finally, in the Kullback-Leibler divergence measure of formula (\ref{modified_log-likelihood_Bekk}) we compute the difference between matrices $\widehat{\Sigma}$ and $H_t$. The modified log-likelihood function is:

\begin{equation}\label{modified_log-likelihood_Bekk}
\begin{array}{ll}
L(\theta)&=-\frac{TN}{2}\log(2\pi)-\frac{1}{2}\sum\limits_{t=1}^T \big(\log|H_t|+ \epsilon'_t H_{t}^{-1} \epsilon_t \big) - KL(\widehat{\Sigma},H_t)\\
\\
&=-\frac{TN}{2}\log(2\pi)-\frac{1}{2}\sum\limits_{t=1}^T \big(\log|H_t|+ \epsilon'_t H_{t}^{-1} \epsilon_t \big) - \frac{1}{2}\sum\limits_{t=1}^T\big( \log(\frac{|H_t|}{|\widehat{\Sigma}|}) + Tr(H_t^{-1}\widehat{\Sigma}) - N\big)\\
\\
&=-\frac{1}{2}\big[N(T\log(2\pi))+\sum\limits_{t=1}^T\big(2\log(|H_t|) + \epsilon'_t H_{t}^{-1} \epsilon_t - \log(|\widehat{\Sigma}|) + Tr(H_t^{-1}\widehat{\Sigma})\big) \big].
\end{array}
\end{equation}

\noindent Observe that the minimization of the distance between the target unconditional covariance matrix $\widehat{\Sigma}$ and $H_t$ forces the values $\sigma_{ij,t}$ to be as close as possible to the values $\sigma_{ij}$, that is, forces the values $\rho_{ij,t}$, $i\ne j$, to be as close as possible to the values $\rho_{ij}$.

\subsection{The DCC model}\label{DCC}

The idea of this model, introduced and analyzed in \cite{EngShep_2001}, is that the conditional covariance matrix $\bf{H_t}$ can be decomposed into the conditional standard deviations $\bf{D_t}$ of each of the $N$ series and a conditional correlation matrix of the returns $\bf{R_t}$. The dynamic of the model is described by the equations (\ref{GeneralGARCH}) and:

\begin{equation}\label{DCC_dynamics}
\bf{H_t}=\bf{D_t}\bf{R_t}\bf{D_t}.
\end{equation}

\noindent The matrix $\bf{D_t}$ is a diagonal matrix and consists of the $N$ univariate GARCH models. Since it is a diagonal non negative matrix with all diagonal elements positive, $\bf{D_t}$ is positive definite. To ensure $\bf{H_t}$ to be positive definite, it is, therefore, necessary that the matrix $\bf{R_t}$ be positive definite with the additional constraint that all its elements have to be equal to or less than 1 by definition. The dynamic of correlation matrix $\bf{R_t}$ is, in turn, derived from another matrix $\bf{Q_t}$ of the form:

\begin{equation}\label{Rt_dynamics}
\bf{R_t}=\bf{\bar{Q}_t^{-1}Q_t\bar{Q}_t^{-1}},
\end{equation}

\noindent were $\bf{\bar{Q}_t}=diag(\bf{Q_t})$ with $diag(\cdot)$ the diagonal of a square matrix. The form of $\bf{Q_t}$ determines the dynamic of the model and its complexity (see, e.g., \cite{Engle_2002, TseTsui_2002}). For example, following \cite{Engle_2002}, the dynamics is:

\begin{equation}\label{Qt_dynamics}
\bf{Q_t}=(1-\theta_1-\theta_2)\bf{\hat{Q}_t} + \theta_1(\epsilon_{t-1}\epsilon'_{t-1}) + \theta_2\bf{Q_{t-1}},
\end{equation}

\noindent with $\bf{\hat{Q}_t}$=$Cov[\epsilon_{t}\epsilon'_{t}]=E[\epsilon_{t}\epsilon'_{t}]$. In order to ensure $\bf{R_t}$ be positive definite, the parameters $\theta_1$ and $\theta_2$ must satisfy:

$$\theta_1\geq 0,\; \theta_2\geq 0,\; \theta_1+\theta_2 < 1.$$
\vskip 8 pt

\noindent The parameter estimation phase is rather difficult, and hence for the DCC model a two stage estimation procedure is provided. In the first stage the parameters of the univariate GARCH models are estimated for each asset series. In the second stage, a second set of parameters are estimated given the parameters found in the previous phase. Referring to the dynamics described in (\ref{Qt_dynamics}) and assuming multivariate Gaussian distributed errors, after the first step only the parameters $\theta_1$ and $\theta_2$ are unknown so they are estimated in the second stage. In this second phase the log-likelihood function is:

\begin{equation}\label{log-likelihood_DCC_II}
L(\theta)=-\frac{1}{2}\sum\limits_{t=1}^T \big(\log(|R_t|)+ \epsilon'_t R_{t}^{-1} \epsilon_t \big).
\end{equation}

\noindent In our approach we are interested in the second stage of the process where the log-likelihood function takes into account the correlation matrix of the assets returns at time $t$. Hence, under the assumption $w_{ij}=\rho_{ij}$, $i\ne j$, given $\delta$, we consider the matrix $\widehat{Z}$ and the matrix $\bf{R_t}$. The corresponding modified log-likelihood function is:

\begin{equation}\label{modified_log-likelihood_DCC}
\begin{array}{ll}
L(\theta)&=-\frac{1}{2}\sum\limits_{t=1}^T \big(\log(|R_t|)+ \epsilon'_t R_{t}^{-1} \epsilon_t \big) - KL(\widehat{Z},R_t)\\
\\
&=-\frac{1}{2}\big[\sum\limits_{t=1}^T\big(2\log(|R_t|) + \epsilon'_t R_{t}^{-1} \epsilon_t - \log(|\widehat{Z}|) + Tr(R_t^{-1}\widehat{Z}) - N \big) \big].
\end{array}
\end{equation}

\noindent As for the modified BEKK model, to highlight the strong relationships among groups of assets in the market, the minimization of the distance between the (target) matrix $\widehat{Z}$ and $\bf{R_t}$ forces the values $\rho_{ij,t}$, $i\ne j$, to be as close as possible to the values $\rho_{ij}$.
\vskip 8 pt

\noindent To conclude this section, we observe that if we set $\delta=0$ and under the assumption that $w_{ij}=\rho_{ij}$, $i\ne j$, the correlation matrix $\widehat{Z}$, and consequently the unconditional covariance matrix $\widehat{\Sigma}$, correspond exactly to the long-run correlation and covariance matrices used as target matrices in \cite{CapMcA_2012}. Hence, on the one hand, our method can be precisely considered as an alternative targeting method w.r.t. the one proposed in \cite{CapMcA_2012}. On the other hand, assuming $\delta>0$, with our method we are, in fact, ``forcing'' our modified MGARCH models in highlighting clusters of highly correlated assets in the simulated series.

\section{Data analysis and test}\label{Data_analysis}

In order to show the effectiveness of our new estimation approach, which aims at correctly identifying highly (positive or negative) correlated groups of assets, given a financial dataset we have to perform a first pre-processing phase by a method of clustering financial time series.
In a broad context, cluster analysis is about assembling a collection of objects, and each of these objects is called a cluster, in which everything should be similar but still as different as possible from everything else in that cluster (see \cite{Sar2017}). Time series clustering creates groupings based on similarities between time series without using pre-existing categorization labels \cite{RadLat2017}. Various approaches were proposed in time series clustering, and various authors attempted to synthesize the literature on the field (see, e.g., \cite{Aghal2015} \cite{Lia2005} \cite{MonVil2014}).
\\
According to \cite{MonVil2014}, there is a wide range of applications for time series clustering that can be found in a variety of domains (e.g., economic or medical time series analysis, visual processing, anomalous activity detection). Different clusters of time series can be formed based on evolving patterns or real-world interactions (group membership can be interpreted as temporal context). In this case, the prediction may also differ from the results of cluster analysis \cite{MahInd1999}.
\\
Time series clustering is particularly relevant in finance. There is a particular interest in discovering typical dynamics of financial markets and the impact of different shocks on time series and portfolio allocation groups. In this respect, according to \cite{PicAl2011} clustering of financial time series is essential to determine how much wealth should be allocated to financial assets and opportunities. Therefore, financial time series need to be clustered to select an appropriate portfolio and analyze an economic system. When studying highly complicated phenomena such as financial time series, one has to deal with substantial heterogeneity and peculiar characteristics and features (see \cite{Sew2011} and \cite{ShiAl2021}). Therefore, robustness measures should address this challenge \cite{DurAl2016}. Moreover, studying their behavior over time in a dynamic framework is also essential, as these systems are associated with uncertainty and for this reason, other approaches representing this uncertainty were proposed in \cite{DraAl2013}.
\\
In this section we provide a statistical framework for classifying time series and an example of applying the proposed method to a group of time series. Thus, our whole approach is based on two stages: In the first stage, we explicitly define the different groups of time series using a clustering procedure, which is helpful to identify the different partitions (groups). In the second stage we apply our new estimation method to correctly recognize the groups obtained so far. In our example, we performs the first stage on the following set of stocks:  Facebook, Apple, Google, Boeing, Microsoft, Amazon, General Motors, Goldman Sachs, JPMorgan Chase, Intel, Verizon Communications, Visa, Cisco, Coca-Cola Company, and Salesforce. They were selected without any particular pattern from the stocks listed on the NYSE, the New York Stock Exchange. The period considered is from 2020-01-01 to 2021-01-01, and the resulting 253 observations refer to the daily closing prices of each financial time series.
\vskip 8 pt
\noindent Following \cite{MonVil2014}, we have to clarify what is the claim in what two data are similar, and in what way we can obtain a good similarity and dissimilarity measure between two different observations (in this case, two different financial time series). These are the central questions of cluster analysis. In this regard, we have explicitly considered a hierarchical clustering algorithm that uses a Pearson correlation-based distance and the \emph{complete} method (see \cite{EveAl2001}) as a helpful approach to distinguish the different groups that we can identify. Concerning financial time series, Pearson's correlation coefficient is widely used in the related financial literature to quantify the degree of similarity or dissimilarity between two time series \cite{PicAl2011}. If we denote ${y}_{t}$ and ${x}_{t}$ as two different time series, we can consider their correlations \cite{MonVil2014}:

\begin{equation}
\operatorname{COR}\left(x_{t}, y_{t}\right)=1- \frac{\sum_{t=1}^{T}\left(y_{t}-\overline{x}_{t}\right)\left(y_{t}-\overline{y}_{t}\right)}{\sqrt{\sum_{t=1}^{t}\left(x_{t}-\overline{x}_{t}\right)^{2}} \sqrt{\sum_{t=1}^{t}\left(y_{t}-\overline{y}_{t}\right)^{2}}},
\end{equation}

\noindent So we can compute the following distance, and we have:

\begin{equation}
d_{Cor}\left(x_{t},y_{t}\right)=1- COR \left(x_{t}, y_{t}\right)
\end{equation}

\begin{figure}[ht]
\begin{center}
\includegraphics[width=0.9\textwidth]{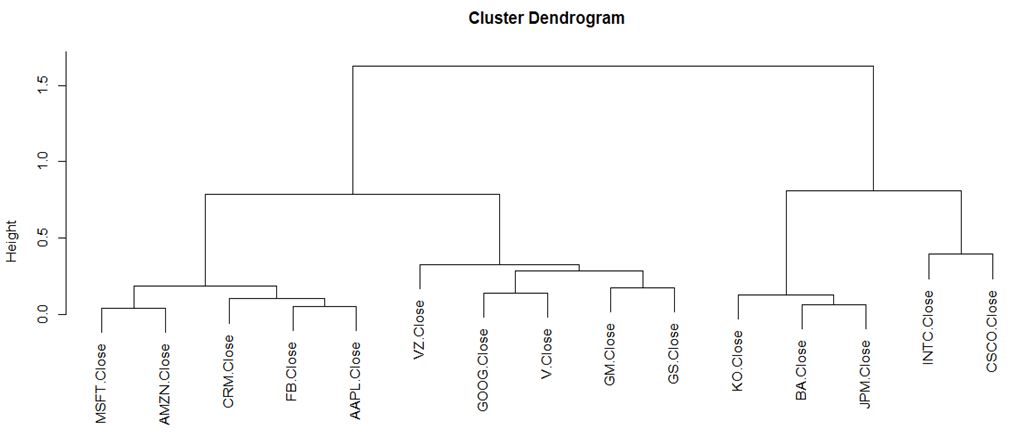}
\end{center}\caption{Clusters obtained in the first phase.}\label{Fig1}
\end{figure}

We apply the distance and cluster approach to develop the dendrogram and interpret the different clusters (see Figure \ref{Fig1}). Then, we follow an exploratory approach to identify the groups of different stocks that are included in each cluster. We use the dendrogram to collect information about the variability of a cluster structure and the overall groups structure.

\noindent No prior knowledge is required for this approach, and it allows understanding patterns in financial time series data without the need for additional information \cite{AhnLee2018}. This approach can be summarized as follows: use similarity measures to identify subgroups (see \cite{WanAl2006}); determine which generic grouping each individual belongs to.
Through the above clustering procedure, we can identify four general clusters (see Figure \ref{Fig1}). First, we can identify Microsoft's most significant joint movement (MSFT) and Amazon (AMZN). At the same time, Facebook (FB) and Apple (AAPL), and finally, Salesforce (CRM) belong to the same cluster. This cluster is clearly based on the technological innovations that can have an overarching impact on different sectors and companies (see also \cite{DraAl2013}). Finally, at the same time, Coca Cola (KO) and, more importantly, Boeing (BA) and JPMorgan Chase (JPM) are in a different cluster.
Furthermore, Intel (INTC) and Cisco (CSCO) can also be considered part of the same cluster. We can further observe the critical role of economic similarities for these two stocks in this case. In this respect, it is possible to suppose relevant economic shocks that affect both financial time series. Finally, in the third cluster, we can observe a slightly different behavior for Verizon Communications (VZ), but also daily solid joint movements for Google (GOOG) and Visa (V), as well as for General Motors (GM) and Goldman Sachs (GS). Overall, through observation and visual analysis of the dendrogram, the distinct clusters obtained will be used as a reference for the evaluation of the effectiveness of our new estimation approach (second stage) described in the following.

\subsection{Estimation phase}\label{Test}

In this section we present some examples of different sizes pointing out the differences between the simulated series obtained by the BEKK model and its modified version, as well as, the DCC model and its modified version. All MGARCH models have been implemented in Matlab R2018 version using the MFE Toolbox code repository by Kevin Sheppard \cite{Sheppard_2013}. In the resulting correlation graphs we detected all the maximal cliques by the Bron-Kerbosch algorithm \cite{BronKer_1973}. The experiments were carried out on a PC equipped with an Intel Core i7-3632MQ processor with 2.20Ghz.

\vskip 8 pt
\noindent \textbf{Example 1}\par

\noindent In this first example we consider only the group of technological innovation assets formed by MSFT, AMZN, CRM, FB and AAPL (see Figure \ref{Fig1}). We compute the 252 log-return values and the corresponding unconditional correlation matrix, which is (see Table \ref{Tab:Example1}):

\begin{table}[ht]
\begin{center}
\begin{tabular}{|c|c|c|c|c|c|}
  \hline
 & \textbf{MSFT}&  \textbf{AMZN} & \textbf{CRM} & \textbf{FB} & \textbf{AAPL}\\
 \hline
  \textbf{MSFT} & 1 & 0.7712 & 0.7690 & 0.6855 & 0.6914\\
  \textbf{AMZN} & 0.7712 & 1 & 0.8435 & 0.7028 & 0.6458\\
  \textbf{CRM}  & 0.7690 & 0.8435 & 1 & 0.7413 & 0.7508\\
  \textbf{FB}   & 0.6855 & 0.7028 & 0.7413 & 1 & 0.6121\\
  \textbf{AAPL} & 0.6914 & 0.6458 & 0.7508 & 0.6121 & 1\\
  \hline
  \end{tabular}
  \end{center}
\caption{The correlation matrix for $N=5$ assets.}
\label{Tab:Example1}
\end{table}

\begin{figure}[ht]
\begin{center}
\includegraphics[width=0.4\textwidth]{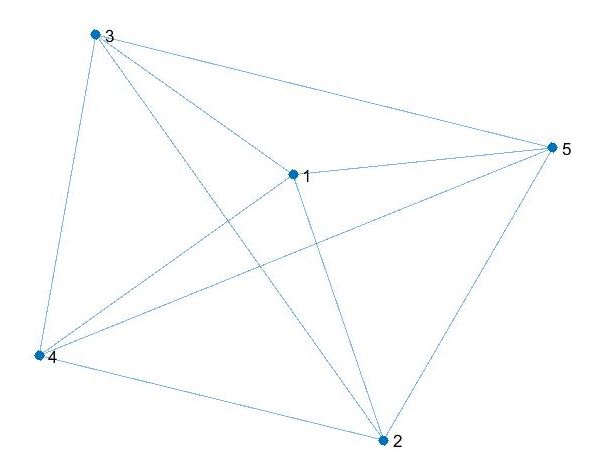}
\end{center}\caption{The graph $G(\delta)$ with $\delta=0.5$.}\label{Fig2}
\end{figure}

\vskip 8 pt
\noindent Let us now consider the (sub)graph $G(\delta)$ of the complete correlation graph $G$ whose vertex set contains only the subset of pairs of assets such that $|\rho_{ij}|>\delta$, with $\delta=0.5$ (see Figure \ref{Fig2}). In the figure, the numbers associated with the vertices correspond to the following labeling: 1-MSFT, 2-AMZN, 3-CRM, 4-FB, 5-AAPL.

\noindent Note that the graph $G(\delta)$ corresponds to the complete graph $G$ since for all pairs of assets $i$ and $j$ we have $\rho_{ij}>0.5$.
\vskip 8 pt
\noindent Consider the whole original dataset consisting of $T=252$ return observations, we first estimates the parameters of the BEKK and DCC models and then the parameters of each of the modified versions of the two models using the log-likelihood functions (\ref{modified_log-likelihood_Bekk}) and (\ref{modified_log-likelihood_DCC}), respectively. Then, using these sets of parameters we simulate (new) $T=252$ log-return observations of the five assets. Our aim is to verify whether our modified models allow to better capture the strong relationships between assets in the stock market. In other words, we want to compare the graph $G(\delta)$ with the simulated graphs $G_S(\delta)$.

\begin{figure}[p]
  \subfloat[The graph $G_S(\delta)$ resulting from the BEKK model]{
	\begin{minipage}[c][1\width]{
	   0.49\textwidth}
	   \centering
	   \includegraphics[width=0.7\textwidth]{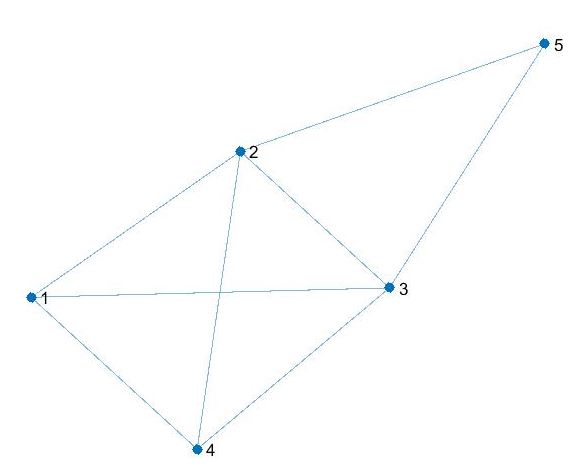}
	\end{minipage}}
 \hfill 	
  \subfloat[The graph $G_S(\delta)$ resulting from the modified BEKK model]{
	\begin{minipage}[c][1\width]{
	   0.49\textwidth}
	   \centering
	   \includegraphics[width=0.7\textwidth]{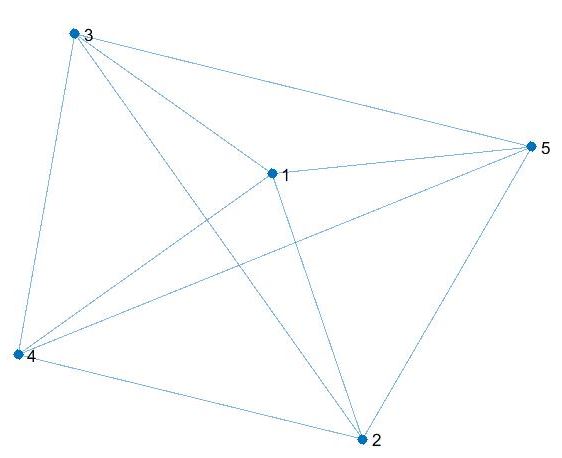}
	\end{minipage}}
\hfill
  \subfloat[The graph $G_S(\delta)$ resulting from the DCC model]{
	\begin{minipage}[c][1\width]{
	   0.49\textwidth}
	   \centering
	   \includegraphics[width=0.7\textwidth]{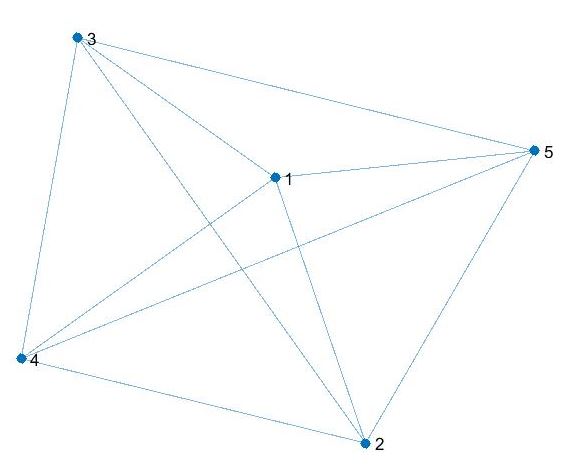}
	\end{minipage}}
 \hfill	
  \subfloat[The graph $G_S(\delta)$ resulting from the modified DCC model]{
	\begin{minipage}[c][1\width]{
	   0.49\textwidth}
	   \centering
	   \includegraphics[width=0.7\textwidth]{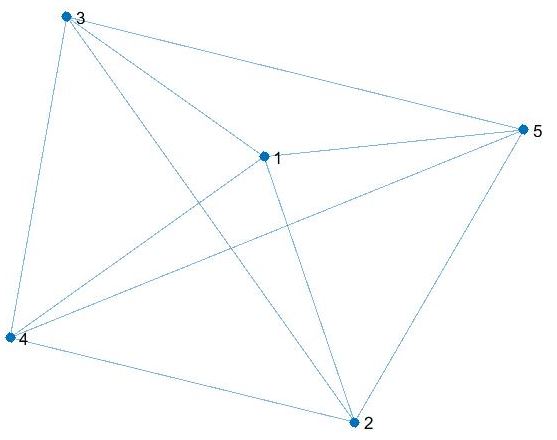}
	\end{minipage}}
\caption{The simulated graphs $G_S(\delta)$ with $\delta=0.5$.}\label{Fig3}
\end{figure}

\noindent Comparing the five graphs, only the original BEKK model was not able to correctly reproduce the original series. In fact, vertex 5, corresponding to asset AAPL, is not connected to all the other vertices despite its correlation value with other assets is greater that $0.5$. In addition to the graphs $G_S(\delta)$, in Table \ref{Tab2:Example1} we report the Frobenius distance between the fitted covariance matrices $H_t$ with respect to the unconditional covariance matrix $\widehat{\Sigma}$, and the Kullback-Leibler divergence between the matrix $\widehat{\Sigma}$ and the unconditional covariance matrix of the new simulated series of log-returns, denoted by $\widehat{\Sigma}_S$, w.r.t. all the considered models. The Frobenius norm is:

\begin{equation}\label{Frobenius}
F=\sqrt{\sum\limits_{t=1}^T Tr[(H_t-\widehat{\Sigma})'(H_t-\widehat{\Sigma})]},
\end{equation}

\noindent and the Kullback-Leibler divergence is:

\begin{equation}\label{KL_distance_simulated}
KL=\frac{1}{2}\big[ \log(\frac{|\widehat{\Sigma}_S|}{|\widehat{\Sigma}|}) + Tr(\widehat{\Sigma}_S^{-1}\widehat{\Sigma}) - N\big].
\end{equation}

\noindent

\begin{table}[ht]
\begin{center}
\begin{tabular}{|c|c|c|c|c|}
  \hline
 & \textbf{BEKK}&  \textbf{modified BEKK} & \textbf{DCC} & \textbf{modified DCC}\\
 \hline
 \hline
  F  & 0.0669 & 0.0555 & 0.0881 & 0.0870\\
  \hline
  KL & 0.0855 & 0.0229 & 0.0162 & 0.0154\\
  \hline
  \end{tabular}
  \end{center}
\caption{Values of the Frobenius and the Kullback-Leibler loss functions: case $N=5$ and $\delta=0.5$}
\label{Tab2:Example1}
\end{table}

\noindent Both functions $F$ and $KL$ measure loss, so that lower values are preferable. We note that the values of the modified MGARCH models are better that the corresponding original models. Here we are not interested in comparing the values of the two loss functions among all the models since, again, it is not our goal to find a single winner.
\vskip 8 pt
\noindent Assume now we are interested in detecting groups of assets with a higher value of correlation. On the basis of the values in Table \ref{Tab:Example1}, we set $\delta=0.71$. The corresponding graph $G(\delta)$ is:

\begin{figure}[ht]
\begin{center}
\includegraphics[width=0.5\textwidth]{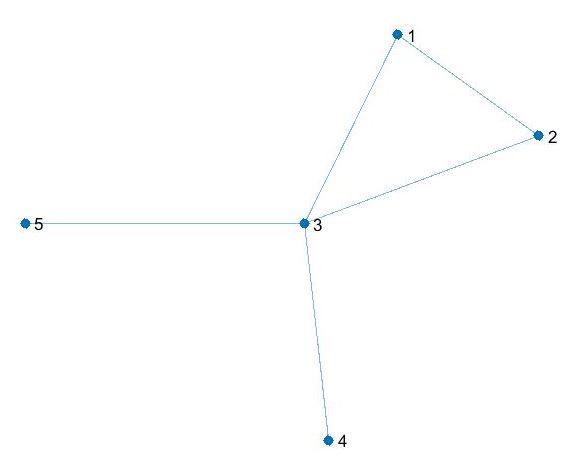}
\end{center}\caption{The graph $G(\delta)$ with $\delta=0.71$.}\label{Fig4}
\end{figure}

\noindent We observe that with this value of $\delta$, only assets $\{1,2,3\}$ form a clique. The goal is to find simulated graphs $G_S(\delta)$, with $\delta=0.71$, as similar as possible to the subgraph $G(\delta)$ of Figure \ref{Fig4}. The results are reported in the following Figure \ref{Fig5} along with the corresponding table (see Table \ref{Tab3:Example1}) reporting the values of the two loss functions considered.

\begin{figure}[p]
  \subfloat[The graph $G_S(\delta)$ resulting from the BEKK model]{
	\begin{minipage}[c][1\width]{
	   0.49\textwidth}
	   \centering
	   \includegraphics[width=0.75\textwidth]{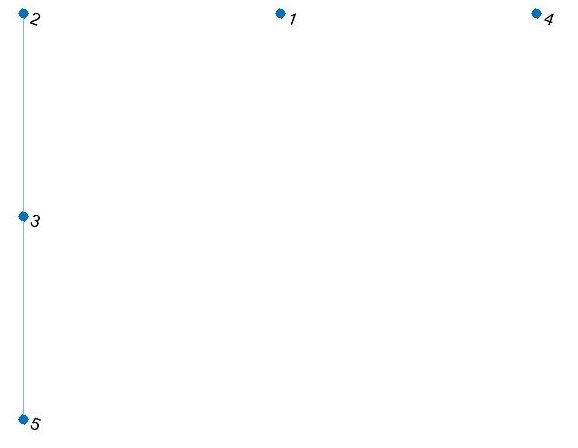}
	\end{minipage}}
 \hfill 	
  \subfloat[The graph $G_S(\delta)$ resulting from the modified BEKK model]{
	\begin{minipage}[c][1\width]{
	   0.49\textwidth}
	   \centering
	   \includegraphics[width=0.75\textwidth]{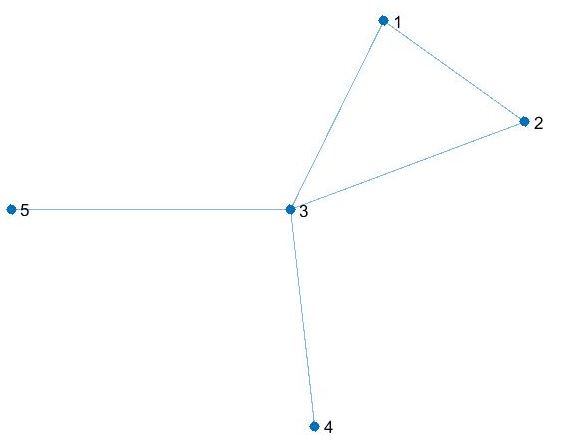}
	\end{minipage}}
\hfill
  \subfloat[The graph $G_S(\delta)$ resulting from the DCC model]{
	\begin{minipage}[c][1\width]{
	   0.49\textwidth}
	   \centering
	   \includegraphics[width=0.75\textwidth]{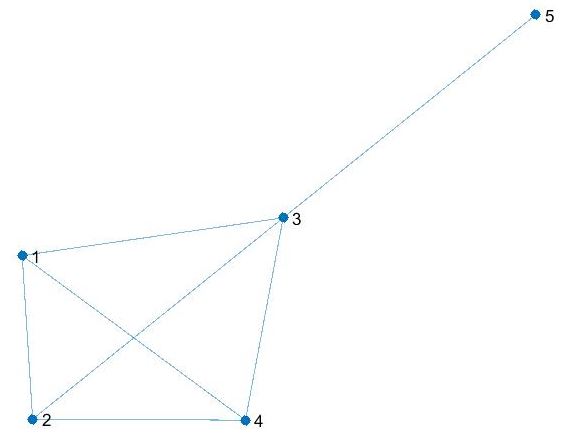}
	\end{minipage}}
 \hfill	
  \subfloat[The graph $G_S(\delta)$ resulting from the modified DCC model]{
	\begin{minipage}[c][1\width]{
	   0.49\textwidth}
	   \centering
	   \includegraphics[width=0.75\textwidth]{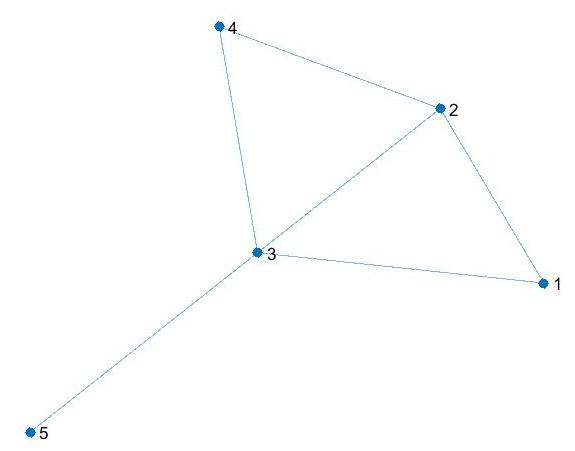}
	\end{minipage}}
\caption{The simulated graphs $G_S(\delta)$ with $\delta=0.71$.}\label{Fig5}
\end{figure}

\noindent The simulated graph of the modified BEKK model exactly replicates $G(\delta)$. The two graphs associated to the DCC and modified DCC models correctly find the clique $C=\{1,2,3\}$, but the DCC model determines a new clique $C'=\{1,2,3,4\}$ that contains $C$, while the modified DCC model introduces the additional edge $(4,2)$.  Thus, these two models ``overestimate'' the correlation between assets pointing out correlations that do not really exist. The original BEKK model has the lowest ability to correctly identify groups of highly correlated assets. On the other hand, the values of the loss functions reported in Table \ref{Tab3:Example1} are in line with the values in Table \ref{Tab2:Example1}.

\begin{table}[ht]
\begin{center}
\begin{tabular}{|c|c|c|c|c|}
  \hline
 & \textbf{BEKK}&  \textbf{modified BEKK} & \textbf{DCC} & \textbf{modified DCC}\\
 \hline
 \hline
  F  & 0.0669 & 0.0575 & 0.0881 & 0.0840\\
  \hline
  KL & 0.0976 & 0.0269 & 0.0306 & 0.0262\\
  \hline
  \end{tabular}
  \end{center}
\caption{Values of the Frobenius and the Kullback-Leibler loss functions: case $N=5$ and $\delta=0.71$}
\label{Tab3:Example1}
\end{table}

\vskip 8 pt
\noindent \textbf{Example 2}
\\
\\
\noindent In this example we consider $N=8$ assets. The set of stocks is formed by the five previously considered assets and the group formed by KO, BA and JPM (see Figure \ref{Fig1}).  In the corresponding graphs the numbers associated with the vertices now correspond to the following labeling: 1-MSFT, 2-AMZN, 3-CRM, 4-FB, 5-AAPL, 6-KO, 7-BA, 8-JPM. We choose this new group of three stocks since the two clusters are adequately separated from each other, so that we expect finding two well separated groups of assets in the following correlation graphs. The unconditional correlation matrix is:

\vskip 8 pt
\begin{table}[ht]
\begin{center}
\begin{tabular}{|c|c|c|c|c|c|c|c|c|}
  \hline
 & \textbf{MSFT}&  \textbf{AMZN} & \textbf{CRM} & \textbf{FB} & \textbf{AAPL}& \textbf{KO} & \textbf{BA} & \textbf{JPM}\\
 \hline
  \textbf{MSFT} & 1 & 0.7712 & 0.7690 & 0.6855 & 0.6914 &0.4038   & 0.4834  &  0.4387 \\
  \textbf{AMZN} & 0.7712 & 1 & 0.8435 & 0.7028 & 0.6458 &0.4810   & 0.5412  &  0.5580 \\
  \textbf{CRM}  & 0.7690 & 0.8435 & 1 & 0.7413 & 0.7508 &0.4798   & 0.5912  &  0.5838 \\
  \textbf{FB}   & 0.6855 & 0.7028 & 0.7413 & 1 & 0.6121 &0.2451   & 0.2697  &  0.3162 \\
  \textbf{AAPL} & 0.6914 & 0.6458 & 0.7508 & 0.6121 & 1 &0.3995   & 0.4562  &  0.4472 \\
  \textbf{KO}   & 0.4038 & 0.4810 & 0.4798 & 0.2451 & 0.3995& 1   & 0.7211  &  0.6461\\
  \textbf{BA}   & 0.4834 & 0.5412 & 0.5912 & 0.2697 & 0.4562& 0.7211& 1 & 0.7304\\
  \textbf{JPM}  & 0.4387 & 0.5580 & 0.5838 & 0.3162 & 0.4472& 0.6461& 0.7304& 1\\
  \hline
  \end{tabular}
  \end{center}
\caption{The correlation matrix for $N=8$ assets.}
\label{Tab1:Example2}
\end{table}

\noindent Consider $\delta=0.5$, the graph $G(\delta)$ is:

\begin{figure}[ht]
\begin{center}
\includegraphics[width=0.5\textwidth]{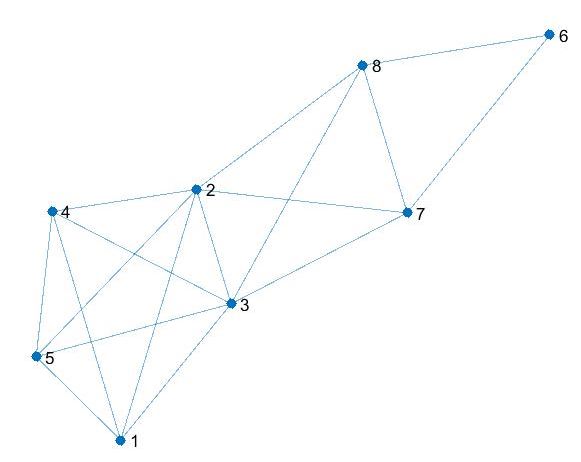}
\end{center}\caption{The graph $G(\delta)$ with $\delta=0.5$ for $N=8$ assets.}\label{Fig6}
\end{figure}

\noindent In $G(\delta)$ we clearly detect the three maximal cliques $C_1=\{1,2,3,4,5\}$, $C_2=\{6,7,8\}$ and $C_3=\{2,3,7,8\}$. In particular, cliques $C_1$ and $C_2$ refer to the two clusters of assets in Figure \ref{Fig1}. The simulated graphs $G_S(\delta)$ are in Figure \ref{Fig7}:

\begin{figure}[p]
  \subfloat[The graph $G_S(\delta)$ resulting from the BEKK model]{
	\begin{minipage}[c][1\width]{
	   0.49\textwidth}
	   \centering
	   \includegraphics[width=0.75\textwidth]{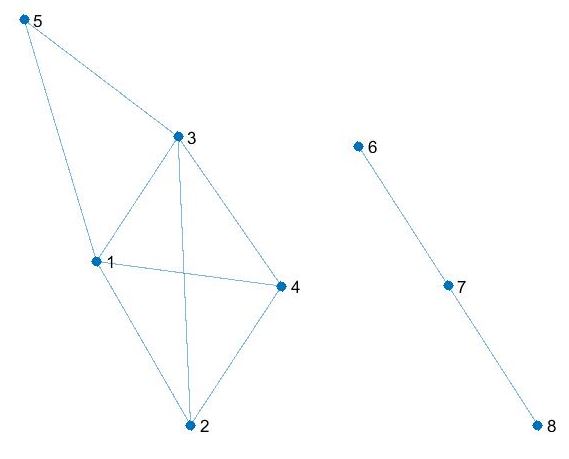}
	\end{minipage}}
 \hfill 	
  \subfloat[The graph $G_S(\delta)$ resulting from the modified BEKK model]{
	\begin{minipage}[c][1\width]{
	   0.49\textwidth}
	   \centering
	   \includegraphics[width=0.75\textwidth]{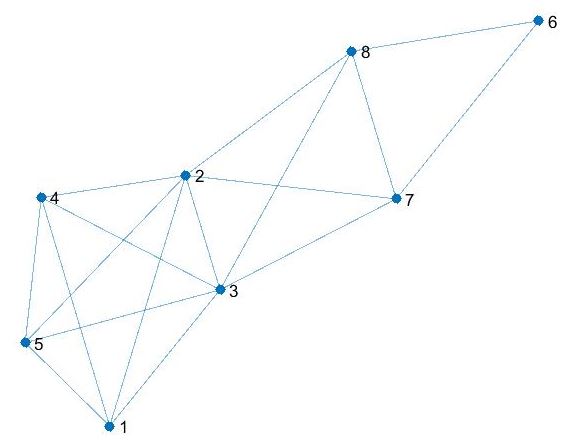}
	\end{minipage}}
\hfill
  \subfloat[The graph $G_S(\delta)$ resulting from the DCC model]{
	\begin{minipage}[c][1\width]{
	   0.49\textwidth}
	   \centering
	   \includegraphics[width=0.75\textwidth]{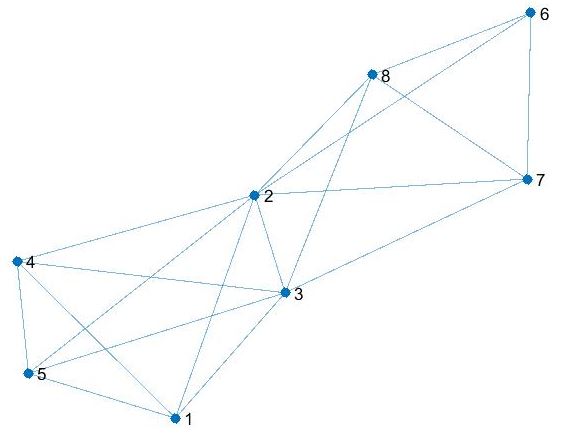}
	\end{minipage}}
 \hfill	
  \subfloat[The graph $G_S(\delta)$ resulting from the modified DCC model]{
	\begin{minipage}[c][1\width]{
	   0.49\textwidth}
	   \centering
	   \includegraphics[width=0.75\textwidth]{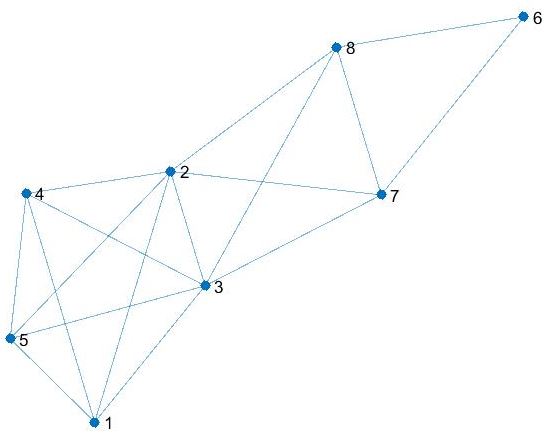}
	\end{minipage}}
\caption{The simulated graphs $G_S(\delta)$ with $\delta=0.5$ for $N=8$ assets.}\label{Fig7}
\end{figure}

\noindent In the following table we report the values of the loss functions $F$ and $KL$ and the additional information on the number of maximal cliques in each of the simulated graphs $G_S(\delta)$.

\begin{table}[ht]
\begin{center}
\small
\begin{tabular}{|c|c|c|c|c|}
  \hline
 & \textbf{BEKK}&  \textbf{modified BEKK} & \textbf{DCC} & \textbf{modified DCC}\\
 \hline
 \hline
  F  & 0.0669 & 0.0575 & 0.0881 & 0.0840\\
  \hline
  KL & 0.0976 & 0.0269 & 0.0306 & 0.0262\\
  \hline
  {Maximal Cliques} & \parbox{2cm}{\{1,2,3,4\}, \{1,3,5\}, \{6,7\}, \{7,8\}} & \parbox{2cm}{\{1,2,3,4,5\}, \{2,3,7,8\}, \{6,7,8\}}& \parbox{2cm}{\{1,2,3,4,5\}, \{2,3,7,8\}, \{2,6,7,8\}}& \parbox{2cm}{\{1,2,3,4,5\}, \{2,3,7,8\}, \{6,7,8\}}\\
  \hline
  \end{tabular}
  \end{center}
\caption{Values of the Frobenius, the Kullback-Leibler loss functions and maximal cliques: case $N=8$ and $\delta=0.5$}
\label{Tab2:Example2}
\end{table}

\noindent The graphs $G_S(\delta)$ related to the two modified MGARCH models are exactly the same as the graph $G(\delta)$. It is worth noticing that this does not mean that the correlation matrices of the simulated returns series with the modified BEKK and DCC models are equal to the correlation matrix of the original series, but that both the modified models are able to detect the same (original) clusters of highly correlated assets. Actually, assuming highly (positive) correlated assets remain unaltered over time, by forecasting future values of the series using the results provided by the modified BEKK and/or DCC models, one might expect to simulate returns series that do not alter too much the correlation structure among assets, as well as, the strong relationships among stocks.
This can be useful in a portfolio selection problem from a diversification viewpoint. It is well-known that correlation represents the degree of relationship between the price movements of different assets included in the portfolio for diversified portfolios. Thus, choosing pairs of assets less correlated decreases the portfolio's overall risk.
Consider, for example, the graph $G_S(\delta)$ in Figure \ref{Fig7}-c referred to the original DCC model, which is very similar to the graph $G(\delta)$ (in fact there is just one additional edge in $G_S(\delta)$, that is, edge $(2,6)$). It includes the maximal clique $C=\{2,6,7,8\}$ showing that asset 2 shows now a correlation value with asset 6 greater than $0.5$, and precisely equal to $0.536$, while from Table \ref{Tab1:Example2} we have $\rho_{26}=0.4810$. Hence, the original DCC model inserts a strong correlation between these two assets, which is not truthful, and, without reason, this might prevent choosing assets 2 and 6 in an optimal portfolio. Finally, note that these considerations are difficult to obtain looking only at the values of the loss functions that show that the modified models behave better than the original ones.
\\
\noindent Finally, observe that the graph $G_S(\delta)$, related to the original BEKK model (see Figure \ref{Fig7}-a), reports a very distorted relationship between the assets in the market. Hence, on the basis of the two above examples, we can state that it seems that the BEKK model benefits more effectively from the introduction of the modified likelihood function.


\vskip 8 pt
\noindent \textbf{Example 3}
\\
\\
\noindent In this example we consider all the $N=15$ assets described in Section \ref{Data_analysis}. In the graphs, the new labels associated to each asset are (see Figure \ref{Fig1}): 1-MSFT, 2-AMZN, 3-CRM, 4-FB, 5-AAPL, 6-VZ, 7-GOOG, 8-V, 9-GM, 10-GS, 11-KO, 12-BA, 13-JPM, 14-INTC, 15-CSCO. The corresponding correlation matrix is in Table \ref{Tab1:Example3}:

\begin{sidewaystable}[p]
 \centering
\begin{tabular}{|c|c|c|c|c|c|c|c|c|c|c|c|c|c|c|c|}
  \hline
 & \textbf{MSFT}&  \textbf{AMZN} & \textbf{CRM} & \textbf{FB} & \textbf{AAPL}& \textbf{VZ} & \textbf{GOOG} & \textbf{V}&  \textbf{GM}&  \textbf{GS}&  \textbf{KO}&  \textbf{BA}&  \textbf{JPM}&  \textbf{INTC}&  \textbf{CSCO} \\
 \hline
  \textbf{MSFT} 	&	1	&	0.7712	&	0.7690	&	0.6855	&	0.6914	&	0.8067	&	 0.4671	&	0.5619	&	0.6656	&	 0.4566	&	0.4038	&	0.4834	&	 0.4387	&	0.5493	 &	0.5777	\\
  \textbf{AMZN} 	&	0.7712	&	1	&	0.8435	&	0.7028	&	0.6458	&	0.7581	&	 0.4539	&	0.6225	&	0.7123	&	 0.5523	&	0.4810	&	0.5412	&	 0.5580	&	0.6333	 &	0.6596	\\
  \textbf{CRM}  	&	0.7690	&	0.8435	&	1	&	0.7413	&	0.7508	&	0.8558	&	 0.4849	&	0.6484	&	0.7830	&	 0.6080	&	0.4798	&	0.5912	&	 0.5838	&	0.7187	 &	0.7171	\\
  \textbf{FB}   	&	0.6855	&	0.7028	&	0.7413	&	1	&	0.6121	&	0.6837	&	 0.2288	&	0.3825	&	0.4519	&	 0.3955	&	0.2451	&	0.2697	&	 0.3162	&	0.5065	 &	0.4991	\\
  \textbf{AAPL} 	&	0.6914	&	0.6458	&	0.7508	&	0.6121	&	1	&	0.6846	&	 0.4191	&	0.5118	&	0.6353	&	 0.4143	&	0.3995	&	0.4562	&	 0.4472	&	0.5170	 &	0.5358	\\
  \textbf{VZ}	&	0.8067	&	0.7581	&	0.8558	&	0.6837	&	0.6846	&	1	&	0.5379	 &	0.6644	&	0.7903	&	0.5877	&	 0.5163	&	0.6020	&	 0.6172	&	0.6535	&	0.6838	 \\
  \textbf{GOOG}	&	0.4671	&	0.4539	&	0.4849	&	0.2288	&	0.4191	&	0.5379	&	1	 &	0.7488	&	0.7014	&	0.4135	&	 0.6972	&	0.7590	&	 0.5881	&	0.4913	&	0.5131	 \\
  \textbf{V}	&	0.5619	&	0.6225	&	0.6484	&	0.3825	&	0.5118	&	0.6644	&	 0.7488	&	1	&	0.7604	&	0.5899	&	 0.6963	&	0.8907	&	 0.7001	&	0.5981	&	 0.6526	\\
  \textbf{GM}	&	0.6656	&	0.7123	&	0.7830	&	0.4519	&	0.6353	&	0.7903	&	 0.7014	&	0.7604	&	1	&	0.6280	&	 0.6541	&	0.7773	&	 0.7363	&	0.6404	&	 0.7295	\\
  \textbf{GS}	&	0.4566	&	0.5523	&	0.6080	&	0.3955	&	0.4143	&	0.5877	&	 0.4135	&	0.5899	&	0.6280	&	1	&	 0.4366	&	0.5975	&	 0.7133	&	0.5698	&	 0.6544	\\
  \textbf{KO}   	&	0.4038	&	0.4810	&	0.4798	&	0.2451	&	0.3995	&	0.5163	&	 0.6972	&	0.6963	&	0.6541	&	 0.4366	&	1	&	0.7211	&	 0.6461	&	0.4908	&	 0.5061	\\
  \textbf{BA}   	&	0.4834	&	0.5412	&	0.5912	&	0.2697	&	0.4562	&	0.6020	&	 0.7590	&	0.8907	&	0.7773	&	 0.5975	&	0.7211	&	1	&	 0.7304	&	0.5663	&	 0.6145	\\
  \textbf{JPM}  	&	0.4387	&	0.5580	&	0.5838	&	0.3162	&	0.4472	&	0.6172	&	 0.5881	&	0.7001	&	0.7363	&	 0.7133	&	0.6461	&	0.7304	&	 1	&	0.5542	&	 0.5984	\\
  \textbf{INTC}	&	0.5493	&	0.6333	&	0.7187	&	0.5065	&	0.5170	&	0.6535	&	 0.4913	&	0.5981	&	0.6404	&	 0.5698	&	0.4908	&	0.5663	&	 0.5542	&	1	&	 0.6700	\\
  \textbf{CSCO}	&	0.5777	&	0.6596	&	0.7171	&	0.4991	&	0.5358	&	0.6838	&	 0.5131	&	0.6526	&	0.7295	&	 0.6544	&	0.5061	&	0.6145	&	 0.5984	&	0.6700	 &	1	\\
  \hline
  \end{tabular}
\caption{The correlation matrix for $N=15$ assets.}
\label{Tab1:Example3}
\end{sidewaystable}

\begin{figure}[p]
\begin{center}
\includegraphics[width=0.7\textwidth]{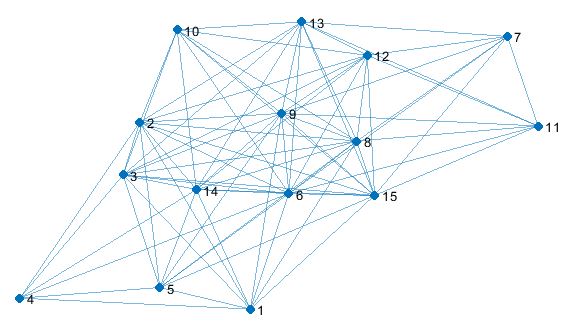}
\end{center}\caption{The graph $G(\delta)$ with $\delta=0.5$ for $N=15$ assets.}\label{Fig8}
\end{figure}

\vskip 8 pt
\noindent The correlation graph for $\delta=0.5$ is depicted in Figure \ref{Fig8}.

\vskip 8 pt
\noindent The maximal cliques in $G(\delta)$ with $\delta=0.5$ are: $C_1=\{1,2,3,4,5,6,14\}$, $C_2=\{1,2,3,5,6,8,9,14,15\}$, $C_3=\{2,3,6,8,9,10,12,13,14,15\}$ and $C_4=\{6,7,8,9,11,12,13,15\}$. These cliques show the most interconnected groups of assets at a correlation level greater than $0.5$. In the following we report the table about the values of the loss functions $F$ and $KL$ and the information on the maximal cliques in each of the simulated graphs $G_S(\delta)$. More precisely, in Table \ref{Tab2:Example3} we report in bold the cliques of the simulated graphs that coincide to the ones of the original graph $G(\delta)$. The modified BEKK model seems to be the best performing one both w.r.t the loss function values and the number of the same maximal cliques of $G(\delta)$. We also note that the first clique in the graph referring to the modified BEKK model has the same number of nodes of clique $C_1$ but vertex 15 in place of vertex 14. We observe that the correlation values of the first six assets in Table \ref{Tab1:Example3} w.r.t. vertices 14 and 15 are very close to each other. Thus, the difference in those two cliques might be due to numerical reasons in the estimation and simulation phase of the modified BEKK model. In any case, all models introduce correlation values, and then a larger number of quite different cliques, that do not show up in the graph $G(\delta)$.

\begin{sidewaystable}[p]
\centering
\small
\begin{tabular}{|c|c|c|c|c|}
  \hline
 & \textbf{BEKK}&  \textbf{modified BEKK} & \textbf{DCC} & \textbf{modified DCC}\\
 \hline
 \hline
  F  & 0.1708 & 0.1682 & 0.1995 & 0.2028\\
  \hline
  KL & 0.4035 & 0.3002 & 0.3309 & 0.3091\\
  \hline
  {Maximal Cliques} & \parbox{4.5cm}{\{1,2,3,4,5,6\}, \{7,8,9,11,12,13\}, \textbf{\{1,2,3,5,6,8,9,14,15\}}, \{1,2,3,6,8,9,10,14,15\}, \textbf{\{2,3,6,8,9,10,12,13,14,15\}}, \{7,8,9,12,13,14,15\}} & \parbox{4.5cm}{\{1,2,3,4,5,6,15\}, \textbf{\{1,2,3,5,6,8,9,14,15\}}, \textbf{\{2,3,6,8,9,10,12,13,14,15\}}, \textbf{\{6,7,8,9,11,12,13,15\}}, \{6,8,9,11,12,13,14,15\}}& \parbox{4.5cm}{\{1,2,3,4,5,6\}, \{1,2,3,5,6,8,9\}, \{2,3,6,8,9,10,15\}, \{3,6,8,9,10,13,15\}, \{3,8,9,10,12,13,15\}, \{3,8,9,10,12,14,15\}, \{7,8,9,11,12,13\}, \{2,8,9,11,14,15\},\{8,9,11,12,13,15\}, \{8,9,11,12,14,15\}}& \parbox{4.5cm}{\{1,2,3,4,6\}, \{1,3,4,5,6\}, \{1,3,5,6,9\}, \{1,2,3,6,8,9,10,13,14,15\}, \{1,3,6,8,9,10,12,13,14,15\}, \{1,6,7,8,9,12,13\}, \{7,8,9,11,12,13\}}\\
  \hline
  \end{tabular}
\caption{Values of the Frobenius, the Kullback-Leibler loss functions and maximal cliques: case $N=15$ and $\delta=0.5$.}
\label{Tab2:Example3}
\end{sidewaystable}

\noindent

\vskip 8 pt
\noindent To evaluate the effectiveness of our method in finding groups of strongly correlated assets, which should be also distinct enough from each other, we consider a higher value for the threshold, that is, $\delta=0.65$. The corresponding correlation graph $G(\delta)$ is in Figure \ref{Fig9}:

\begin{figure}[p]
\begin{center}
\includegraphics[width=0.7\textwidth]{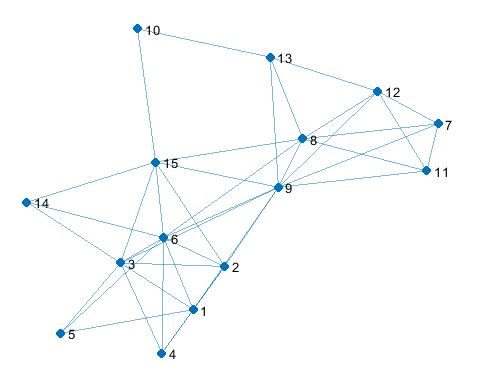}
\end{center}\caption{The graph $G(\delta)$ with $\delta=0.65$ for $N=15$ assets.}\label{Fig9}
\end{figure}

\noindent As expected, the graph $G(\delta)$, with $\delta=0.65$, is sparser than the graph $G(\delta)$, with $\delta=0.5$. In the new graph we identify a larger number of cliques, but of cardinality less than the cardinality of the maximal cliques when $\delta=0.5$.

\begin{sidewaystable}
\centering
\small
\begin{tabular}{|c|c|c|c|c|}
  \hline
 & \textbf{BEKK}&  \textbf{modified BEKK} & \textbf{DCC} & \textbf{modified DCC}\\
 \hline
 \hline
  F  & 0.1708 & 0.1659 & 0.1995 & 0.2038\\
  \hline
  KL & 0.3721 & 0.2982 & 0.2627 & 0.2433\\
  \hline
  {Maximal Cliques} & \parbox{4.5cm}{\textbf{\{1,2,3,4,6\}}, \{1,2,3,5,6\}, \{2,3,5,6,9\}, \{2,3,6,9,14\}, \{3,6,9,14,15\}, \textbf{\{7,8,9,11,12\}}, \{8,9,11,12,13\}, \{6,8,9\}, \{9,10,13\}}
  & \parbox{4.5cm}{\textbf{\{1,2,3,4,6\}}, \textbf{\{1,2,3,6,9\}}, \{1,2,3,5,6\}, \{2,3,6,9,14,15\}, \{3,6,8,9,14,15\}, \textbf{\{7,8,9,11,12\}}, \{8,9,11,12,13\}, \{8,9,12,14\}, \{9,10,12,13\}, \{9,10,15\}}
  & \parbox{4.5cm}{\textbf{\{1,2,3,4,6\}}, \{2,3,6,9\}, \{3,5\}, \{3,14\}, \{7,8,9,12\}, \{7,8,11,12\}, \textbf{\{8,9,12,13\}}, \textbf{\{10,13\}}}
  & \parbox{4.5cm}{\textbf{\{1,2,3,4,6\}}, \{1,2,3,5,6\}, \textbf{\{1,2,3,6,9\}}, \{7,8,9,12\}, \textbf{\{8,9,12,13\}}, \textbf{\{10,13\}}, \{7,8,11,12\}, \{3,15\}}\\
  \hline
  \end{tabular}
\caption{Values of the Frobenius, the Kullback-Leibler loss functions and maximal cliques: case $N=15$ and $\delta=0.65$.}
\label{Tab3:Example3}
\end{sidewaystable}

\vskip 8 pt
\noindent In Table \ref{Tab3:Example3} we report in bold the cliques of the simulated graphs that coincide to the ones in $G(\delta)$, with $\delta=0.65$. It is evident that when $N$ grows, the results are less clear-cut w.r.t. the two previous examples. On the one hand, this can be due that the log-likelihood functions to optimize are highly nonlinear functions and it is hard to find provably optimal solutions. Local solutions may be problematic and this creates difficulties in the estimation of the models. Additionally, since in the optimization phase matrices have to be inverted in each iteration, it makes the overall computation demanding unless $N$ is small. In any case, Tables \ref{Tab2:Example3} and \ref{Tab3:Example3} still highlight that the modified MGARCH models performs slightly better than the original ones, in particular, when $\delta$ increase. For example, for $\delta=0.65$ the simulated series obtained with the modified BEKK and DCC models are still able to better capture the strong relationships among assets in the market than the corresponding original models.

\section{Discussion and conclusions}
\label{Conclusions}

In this paper, we advance a method for improving the estimation phase of financial time series to improve the analysis and evaluation of large portfolios of financial assets whose performance strictly depends on the correlation among assets. Several different estimation models were proposed in the literature; among others, the family of models referred to as multivariate GARCH models are the most widely used. These models take into account that financial time series suffer from heteroscedasticity and that correlation matrices are time-varying. This paper considers two such models, namely the BEKK and DCC models. Furthermore, we modified the log-likelihood objective function based on the idea that in a financial market, there are often specific groups of assets that are highly correlated, and these relationships remain unaltered over time. Hence, in the likelihood function, we introduced a term referring to a loss measure computed on the difference between the time-varying covariance/correlation matrices and the covariance/correlation matrix estimated w.r.t. the whole in-sample period. Given the set of estimated parameters, we use them for simulating new time series to evaluate the effectiveness of the estimation phase. We also propose a new approach for the evaluation phase based on network analysis and, more precisely, on detecting maximal cliques in correlation graphs.

Results from an empirical case study are encouraging, particularly when the number of assets is not large, and we are likely able to find (global) optimal solutions in the maximization of the log-likelihood objective functions. In fact, in these cases, we observed a significant improvement in the ability of the modified MGARCH models to replicate the correlation structure of the assets in the market compared with the original models. When the number of assets increases, the estimation of the models involves somewhat heavy computations because they contain a large number of parameters, and there is no guarantee of finding provably global optimal solutions but only locally optimal solutions. This undermines the ability of the estimated models to replicate the correlation structure of the assets correctly. Despite this, even when the number of assets is large, the modified MGARCH models seem to perform better than the original ones. Overall, our experiments show that the BEKK model is the one that benefits most from the modification of the log-likelihood function.
\vskip 8 pt

\noindent To conclude, we cannot state whether the modified models always prevail over the original ones, and much more experiments are needed. This, therefore, leaves plenty of room for further research. On the one hand, one can experiment with other loss functions introduced in the log-likelihood objective functions to improve the estimates. In fact, we observe that our approach is extremely flexible. Different loss functions can be considered without imposing additional constraints on covariance or correlation matrices.
On the other hand, much attention should be paid in order to improve the optimization phase. This involves developing ad hoc strategies for finding global optimal solutions or solutions close to the optimal ones. In addition, the development of Metaheuristic procedures could be a further line of research to improve the optimization phase. Finally, finding new network indexes that better highlight other peculiar aspects of a financial market is worth investigating.

\newpage

\bibliographystyle{plain}

\end{document}